\input harvmac.tex \input epsf 
\newcount\figno \figno=0
\def\fig#1#2#3{
\par\begingroup\parindent=0pt\leftskip=1cm\rightskip=1cm\parindent=0pt
\baselineskip=11pt \global\advance\figno by 1 \midinsert \epsfxsize=#3
\centerline{\epsfbox{#2}} \vskip 12pt {\bf Figure \the\figno:}  #1\par
\endinsert\endgroup\par } \def\figlabel#1{\xdef#1{\the\figno}}
\def\encadremath#1{\vbox{\hrule\hbox{\vrule\kern8pt\vbox{\kern8pt
\hbox{$\displaystyle #1$}\kern8pt} \kern8pt\vrule}\hrule}} \batchmode
\font\bbbfont=msbm10 \errorstopmode \newif\ifamsf\amsftrue
\ifx\bbbfont\nullfont
\amsffalse \fi \ifamsf \def\IR{\hbox{\bbbfont R}} \def\IZ{\hbox{\bbbfont
Z}}
\def\IF{\hbox{\bbbfont F}} \def\IP{\hbox{\bbbfont P}} \else
\def\IR{\relax{\rm
I\kern-.18em R}} \def\IZ{\relax\ifmmode\hbox{Z\kern-.4em
Z}\else{Z\kern-.4em
Z}\fi} \def\IF{\relax{\rm I\kern-.18em F}} \def\IP{\relax{\rm I\kern-.18em
P}}
\fi

  \def\tilde{\widetilde}
   
\def\tr{{\rm Tr ~}} \font\zfont = cmss10 
\def\ZZ{\hbox{\zfont Z\kern-.4emZ}} 

 %

 

\def\journal#1&#2(#3){\unskip, \sl #1\ \bf #2 \rm(19#3) }
\def\andjournal#1&#2(#3){\sl #1~\bf #2 \rm (19#3) }

\def\frac#1#2{{#1\over#2}}

\def\inbar{\,\vrule height1.5ex width.4pt depth0pt}
\def\IC{\relax\hbox{$\inbar\kern-.3em{\rm C}$}} \def\IR{\relax{\rm
I\kern-.18em
R}} \def\IP{\relax{\rm I\kern-.18em P}}    

\def\slash#1{\mathord{\mathpalette\c@ncel{#1}}} \overfullrule=0pt

\def\ZZ{{\cal Z}}  

\def\underrel#1\over#2{\mathrel{\mathop{\kern\z@#1}\limits_{#2}}}

 \catcode`\@=12


\def\tr{{\rm tr}}


\Title{\vbox{\rightline{hep-th/9712145} \rightline{CERN-TH/97-366}
\rightline{IASSNS--HEP--97/135}
}}
{\vbox{\centerline{Branes and Six-Dimensional Supersymmetric Theories}}}


\centerline{Amihay Hanany} \smallskip{\it
\centerline{School of Natural Sciences} \centerline{Institute for Advanced
Studies} \centerline{Princeton, NJ 08540, USA}} \centerline{\tt
hanany@ias.edu}

\vskip .3cm
\centerline{Alberto Zaffaroni} \smallskip{\it
 \centerline{Theory Division, CERN} \centerline{CH-1211 Geneve 23, Switzerland}} \centerline{\tt
Alberto.Zaffaroni@cern.ch}

\vskip .1in


\noindent
We consider configurations of six-branes, five-branes and eight-branes in various
superstring backgrounds. These configurations give rise to $(0,1)$
supersymmetric theories in six dimensions. The condition for RR charge
conservation of a brane configuration translates to the condition that the
corresponding field theory is anomaly-free. Sets of infinitely many models with non-trivial RG fixed points at strong coupling are
demonstrated. Some of them reproduce and generalise the world-volume theories
of $SO(32)$ and $E_8\times E_8$ small instantons. All the models are shown to be connected by smooth transitions.
In particular, the small instanton transition for which a tensor multiplet is
traded for 29 hypermultiplets is explicitly demonstrated.
The particular limit in which these theories can be considered as six-dimensional string theories without gravity are discussed.
New fixed points (string theories) associated with $E_n$ global symmetries are
discovered by taking the strong string coupling limit. 
\vskip .3truecm
\noindent
CERN-TH/97-366
\Date{December 97}

\lref\am{ P. S. Aspinwall and  D. R. Morrison, {\it Point-like Instantons on K3
Orbifolds},  hep-th/9705104.}

\lref\wpbl{M. Berkooz, R. G. Leigh, J. Polchinski, J. H. Schwarz, N.
       Seiberg and E. Witten, {\it Anomalies, Dualities, and Topology of D=6
N=1 Superstring Vacua},  Nucl. Phys. B475 (1996) 115, hep-th/9605184.}

\lref\hm{ M. R. Douglas and  G. Moore, {\it D-branes, Quivers, and ALE
Instantons}, hep-th/9603167.}

\lref\vo{ H. Ooguri and C. Vafa, {\it Two-Dimensional Black Hole and
Singularities of CY Manifolds}, Nucl. Phys. B463 (1996) 55, hep-th/9511164.}

\lref\gj{E. G. Gimon and C. V. Johnson, {\it  K3 Orientifolds},  Nucl. Phys.
B477 (1996) 715, hep-th/9604129.}

\lref\branes{ J. Polchinski, S. Chaudhuri and  C. V. Johnson, {\it Notes on
D-Branes},  hep-th/9602052.}

\lref\horava{ P. Horava and  E. Witten, {\it Heterotic and Type I String
Dynamics from Eleven Dimensions},  Nucl. Phys. B460 (1996) 506,
hep-th/9510209.}

\lref\pol{ J. Polchinski, {\it Tensors from K3 Orientifolds},
Phys. Rev. D55 (1997) 6423, hep-th/9606165.}

\lref\seibfive{N. Seiberg, {\it  Five-Dimensional SUSY Field Theories,
Non-trivial Fixed Points and String Dynamics}, Phys. Lett. B388 (1996) 753,
hep-th/9608111.}

\lref\sagb{M. Bianchi and  A. Sagnotti, {\it On the Systematic of Open String
Theories}, Phys. Lett. B247 (1990) 517.}
\lref\sagbtwo{M. Bianchi and  A. Sagnotti, {\it Twist Symmetry and Open String
Wilson Lines}, Nucl. Phys. B361 (1991) 519.}
\lref\sagtwo{A. Sagnotti in {\it Cargese 87, Non-Perturbative Quantum Field
Theory}, ed. G. Mack, Pergamon Press NY 1988.}

\lref\kut{S.  Elitzur, A.  Giveon, and D.  Kutasov, {\it Branes and $N=1$
Duality in String Theory}, hep-th/9702014.}


\lref\tatar{R. Tatar {Dualities in 4D Theories with Product Gauge Groups
from
Brane Configurations}, hep-th/9704198.}

\lref\kuttt{S.  Elitzur, A.  Giveon, D.  Kutasov, E.  Rabinovici and A.
Schwimmer, {\it Brane Dynamics and N=1 Supersymmetric Gauge Theory},
hep-th/9704104.}

\lref\jon{N.  Evans, C.  V.  Johnson and A.  D.  Shapere, {\it
Orientifolds,
Branes, and Duality of 4D Gauge Theories}, hep-th/9703210.}

\lref\hw{A.  Hanany and E.  Witten, {\it Type IIB Superstrings, BPS
Monopoles,
and Three-Dimensional Gauge Dynamics}, IASSNS-HEP-96/121, hep-th/9611230.}

\lref\gp{E.  G.  Gimon and J.  Polchinski, {\it Consistency Conditions for
Orientifolds and D-Manifolds}, Phys. Rev.  D54 (1996) 1667, hep-th/9601038.}

\lref\dbl{M.  Berkooz, M.  R.  Douglas, R.  G.  Leigh, {\it Branes
Intersecting
at Angles}, hep-th/9606139, Nucl. Phys.  B480 (1996) 265-278.}

\lref\kleb{ U.  H.  Danielsson, G.  Ferretti and I.  R.  Klebanov, {\it
Creation
of Fundamental Strings by Crossing D-branes}, hep-th/9705084;

O.  Bergman, M. R.  Gaberdiel and G.  Lifschytz, {\it Branes, Orientifolds
and
the Creation of Elementary Strings}, hep-th/9705130.}

\lref\barb{J.  L.  F.  Barbon, {\it
Rotated Branes and $N=1$ Duality}, CERN-TH/97-38, hep-th/9703051.}

\lref\witten{E.  Witten, {\it Solutions of Four-Dimensional Field Theories via
M
 Theory}, hep-th/9703166.}

\lref\brodie{J.H.  Brodie and A.  Hanany, {\it Type IIA Superstrings,
Chiral
Symmetry, and $N=1$ 4D Gauge Theory Duality}, hep-th/9704043}

\lref\dug{M.  R.  Douglas, {\it Branes within Branes}, hep-th/9512077}

\lref\ahup{O.  Aharony and A.  Hanany, unpublished.}

\lref\ah{O.~Aharony and A.~Hanany, {\it Branes, Superpotentials and
Superconformal Fixed Points}, hep-th/9704170.}

\lref\polc{J.  Polchinski, {\it Dirichlet-Branes and Ramond-Ramond
Charges},
Phys.  Rev.  Lett.  75 (1995) 4724}

\lref\alwis{S.  P.  de Alwis, {\it Coupling of branes and normalization of
effective actions in string/M-theory}, hep-th/9705139}

\lref\witt{E.  Witten, {\it String Theory Dynamics In Various Dimensions},
Nucl. Phys.  B443 (1995) 85, hep-th/9503124.}

\lref\polstro{J.  Polchinski and A.  Strominger, {\it New Vacua for Type II
String Theory}, Phys. Lett.  B388 (1996) 736.}

\lref\karch{I.  Brunner and A.  Karch, {\it Branes and Six-Dimensional
Fixed
Points}, hep-th/9705022.}

\lref\barak{B.  Kol, {\it 5d Field Theories and M Theory}, hep-th/9705031.}

\lref\lll{K. Landsteiner, E. Lopez, D. A. Lowe, {\it $N=2$ Supersymmetric
Gauge
Theories, Branes and Orientifolds}, hep-th/9705199.}

\lref\tel{A. Brandhuber, J. Sonnenschein, S. Theisen, S. Yankielowicz,
{\it M Theory And Seiberg-Witten Curves: Orthogonal and Symplectic Groups},
hep-th/9705232.}

\lref\oog{J. de Boer, K. Hori, H. Ooguri, Y. Oz, Z. Yin,
{\it Mirror Symmetry in Three-Dimensional Gauge Theories, SL(2,Z) and
D-Brane
Moduli Spaces}, hep-th/9612131.}

\lref\costas{C. Bachas, M. R. Douglas, M. B. Green, {\it Anomalous Creation
of
Branes}, hep-th/9705074.}

\lref\ted{A. Brandhuber, J. Sonnenschein, S. Theisen, S. Yankielowicz,
{\it Brane Configurations and 4D Field Theory Dualities}, hep-th/9704044.}

\lref\vafa{S. Katz, C. Vafa {\it Geometric Engineering of N=1 Quantum Field
Theories},  hep-th/9611090.}

\lref\vafadue{M. Bershadsky, A. Johansen, T. Pantev, V. Sadov, C. Vafa {\it
F-theory, Geometric Engineering and N=1 Dualities}, hep-th/9612052.}

\lref\vafatre{C. Vafa, B. Zwiebach, {\it N=1 Dualities of SO and USp Gauge
Theories and T-Duality of String Theory}, hep-th/9701015.}

\lref\vafaquattro{H. Ooguri, C. Vafa {\it Geometry of N=1 Dualities in Four
Dimensions} hep-th/9702180.}

\lref\ahn{C. Ahn, K. Oh {\it Geometry, D-Branes and N=1 Duality in Four
Dimensions I}, hep-th/9704061.}

\lref\ahndue{C. Ahn {\it Geometry, D-Branes and N=1 Duality in Four
Dimensions
II}, hep-th/9705004.}

\lref\ahntre{C.Ahn, R. Tatar, {\it Geometry, D-branes and N=1 Duality in
Four
Dimensions with Product Gauge Group}, hep-th/9705106.}

\lref\tatar{R. Tatar, {\it Dualities in 4D Theories with Product Gauge
Groups
from Brane Configurations}, hep-th/9704198.}

\lref\bhoy{J. de Boer, K. Hori, Y. Oz, Z. Yin,
{\it Branes and Mirror Symmetry in $N=2$ Supersymmetric Gauge Theories in
Three Dimensions}, hep-th/9702154.}

\newsec{Introduction}

\lref\seib{N. Seiberg, {\it Non-trivial Fixed Points of The Renormalization
Group in Six Dimensions}, hep-th/9609161; Phys.  Lett.  {\bf B390}
(1996)
753.}

\lref\dani{U.  Danielsson, G Ferretti, J.  Kalkkinen and P.  Stjernberg,
{\it Notes on Supersymmetric Gauge Theories in Five and Six Dimensions},
hep-th/9703098.}

\lref\ken{K. Intriligator, {\it RG Fixed Points in Six
Dimensions via Branes at Orbifold Singularities},  Nucl. Phys. B496 (1997)
177-190, hep-th/9702038.}

\lref\LLnew{K. Landsteiner and E. Lopez, {\it New Curves from Branes}, hep-th/9708118.}
\lref\kenju{J. D. Blum and K. Intriligator, {\it Consistency Conditions for
Branes
at Orbifold Singularities},  hep-th/9705030.}

\lref\kenjutwo{
J. D. Blum and K. Intriligator, {\it New Phases of String Theory and 6d RG
Fixed
Points via Branes at Orbifold Singularities}, hep-th/9705044.}

\lref\karch{I.  Brunner and A.  Karch, {\it Branes and Six-Dimensional
Fixed
Points}, hep-th/9705022.}
 
\lref\vaber{M. Bershadsky and C.  Vafa, {\it  Global Anomalies and Geometric
Engineering of Critical Theories in Six Dimensions}, hep-th/9703167.}  

\lref\GH{O. J. Ganor and  A. Hanany, {\it Small $E_8$ Instantons and Tensionless
Non-Critical Strings}, hep-th/9602120, Nucl. Phys. B474 (1996) 122.}

\lref\SW{N. Seiberg and  E. Witten,
{\it Comments on String Dynamics in Six Dimensions}, hep-th/9603003,
Nucl. Phys. B471 (1996) 121.}

\lref\AH{O. Aharony and A. Hanany,
{\it Branes, Superpotentials and Superconformal Fixed Points}, hep-th/9704170,
Nucl. Phys. B504 (1997) 239.}

By now it is clear that six-dimensional theories may have a much richer
behaviour than expected. The interplay between quantum field theory
considerations and the branes construction of gauge theories led
to a variety of new and surprising information. In the infrared, six-dimensional gauge
theories
flow to free theories. One of the crucial inputs from
string
theory is that we can find theories which are at an interacting fixed point
at strong coupling \seib.
\lref\sagnotti{A.Sagnotti, {\it A Note on the Green - Schwarz Mechanism in Open - String Theories}, Phys. Lett. B294 (1992) 196, hep-th/9210127.}

$(0,1)$ theories in 6d arise
naturally on the world-volume of a D5-brane in the Type I theory, which
represents a small $SO(32)$ instanton. The $SU(2)$ gauge theory with $16$
fundamentals living in the world-volume is IR-free, as any gauge theory
in six dimensions. More exotic theories, which contain tensionless strings,
arise in the case of an $E_8$ small instanton of the heterotic strings.
\GH,\SW.
These
theories are believed to be interacting local quantum field theories at a
non-trivial fixed point. General conditions for the existence of 
non-trivial
strong coupling  fixed points were discussed in \refs{\seib,\dani,\vaber}.
Tensor multiplets are usually needed to cancel gauge anomalies\sagnotti.

The theory of small instantons both of $SO(32)$ and $E_8\times E_8$ is a
good
laboratory for finding non-trivial examples of six-dimensional gauge theories.
In Type I (or heterotic) theory $k$ $SO(32)$ instantons of zero size are
described by $k$ D5-branes which have a $USp(2k)$ gauge theory 
with 16 fundamentals and an antisymmetric living in their world-volume. On the other hand, the
theory of small $E_8\times E_8$ instantons in the heterotic string is a
more
exotic theory, with tensor multiplets and tensionless strings. There is a
Coulomb
branch  parametrized by the scalar partner of the tensor, and a transition to a Higgs branch by higgsing the tensor
multiplet
to 29 hypermultiplets can take place. Because of the self-duality of the tensor
field, this transition is far outside the perturbative region, but the theory
is
believed to be an interacting theory at a non-trivial RG fixed point.
In \refs{\ken\kenju\kenjutwo}, it was discovered that even $SO(32)$ instantons
can have such an
exotic behaviour when they live on top of an orbifold singularity, and a
large
class of gauge theories describing these instantons was determined. These
theories are other good candidates for non-trivial fixed points. Other
theories
were found in \am\ by considering small $E_8\times E_8$ instantons living on
top of orbifold singularities.

\lref\wittennew{E. Witten, {\it  New ``Gauge'' Theories In Six Dimensions},
hep-th/9710065.}

\lref\seiberg{N. Seiberg, {\it Matrix Description of M-theory on $T^5$ and
$T^5/Z_2$}, hep-th/9705221.}

\lref\intr{K. Intriligator, {\it New String Theories in Six Dimensions via
Branes at Orbifold Singularities}, hep-th/9708117.}

\lref\hz{A. Hanany and A. Zaffaroni, {\it Chiral Symmetry from Type IIA
Branes},
hep-th/9706047.}

All these theories, being unrenormalizable, need more data in the ultraviolet
to
be defined. String theory is supposed to provide the necessary
ultraviolet data. It was recently pointed out \seiberg\ that a new class of
theories in six dimensions, without gravity but with stringy excitations,
can be realized on the world-volume of five-branes in the limit
$g_s\rightarrow 0$ with $M_s$ fixed. After compactification, these theories
inherit the T-duality of the ten-dimensional string theory.
For generalisations, see \wittennew.
All the theories we discussed before appear as the limit of some of these
six-dimensional string theories \intr, which can be used to provide their
ultraviolet definition. Having a kind of T-duality, the full six-dimensional
theory is not a local quantum field theory.

In this paper we provide a realization of six-dimensional gauge theories
using branes (in the spirit of \hw), which contains and generalizes the examples
in \ken\ and \am. We provide new examples of six-dimensional theories with non-trivial fixed points and
a general framework for studying their properties. The existence of these
theories could have been guessed by the fact that they are anomaly-free. Their
explicit construction as a system of branes in string theory
guarantees that they indeed exist. Using string duality for representing them as heterotic $SO(32)$ or $E_8\times E_8$ instantons, it can moreover be
shown \refs{\seiberg,\intr}, for most of them, that gravity decouples,  giving consistent six-dimensional string theories.  

\lref\wp{J. Polchinski and  E. Witten, {\it  Evidence for Heterotic - Type I String Duality}, Nucl. Phys. B460 (1996) 525, hep-th/9510169.}

Surprisingly enough, the low energy theory describing $SO(32)$ and $E_8\times E_8$ instantons appears on the same footing, as particular cases of the same brane
construction. The interpretation as instantons becomes manifest only when
we consider the system as embedded in a particular Type I$'$ string theory background. The interpretation of the system as describing small $SO(32)$ instantons is appropriate for
a generic configuration of background branes in the Type I$'$ theory at weak coupling. The global symmetry of such theories is in general a subgroup of
$SO(32)$. Following \refs{\wp,\seibfive}, we can find theories in which the
global symmetry is enhanced (at the fixed point) to exceptional groups. This involves choosing a Type I$'$ background in which the coupling constant diverges at
the orientifold points. We find that, in the strong coupling limit, the theories with enhanced global symmetry have a natural interpretation as $E_8\times E_8$ small instantons. 

The anomaly cancellation is derived as a tree-level charge conservation condition for the spacetime fields, which can be read off directly from the equations of motion. This must be contrasted with the more complicated tadpole computation in \ken\ and the geometrical analysis in \am. As a product of the study of the six-dimensional theories, we give a T-dual
representation of the $SO(32)$ instantons studied in \refs{\ken,\kenju,\kenjutwo}, in which most of the subtleties involved in the subject have a natural and
simple interpretation. We also provide a large class of theories with exceptional global symmetry, which we propose as low energy theories describing
small instantons that partially break $E_8\times E_8$. 

The realization of six-dimensional gauge theories using branes
was provided in \karch\ in the spirit of \hw, and further studied in \hz.
We start in section 2 and 3 by reviewing this construction, we give the
ingredients to
construct general gauge theories and we show how the charge conservation
argument of \hz\
gives the right anomaly cancellation conditions.
 We derive the explicit connection
with the models in \ken\ and \am , which are both contained as particular cases
in the brane construction. The interpretation of the theories in \ken\ and \am\
as small instantons in the  $SO(32)$ and $E_8\times E_8$ heterotic string,
respectively, is not manifest in this general set-up, but can be recovered by
exploring T- and strong coupling dualities of the model. In section 4, we
perform a
T-duality that maps  the brane configuration to the system of D5-branes at
orbifold singularities studied in \ken. In section 5, by exploring the strong
coupling limit of the Type I$'$ theory in which the brane system is embedded, we
recover the theory of $E_8\times E_8$ instantons. A brane configuration
realizing
the theory in \am\ was already proposed in the context of the three-dimensional
mirror symmetry \hw. We show that the same configuration, when interpreted as a
six-dimensional gauge theory, is actually the low energy theory of $E_8\times
E_8$ instantons. We are also able to generalize the theory in \am\ to the case
in which the $E_8\times E_8$ spacetime group is broken to a subgroup by the
small instantons. In section 6, we discuss the decoupling of gravity in the
models discussed in this paper.
At the final stages of this project, we were informed of a paper by Brunner
and Karch, which has some overlap with the material in the present work.

\newsec{Six-dimensional gauge theories}

Here we give the general set-up for constructing six-dimensional gauge
theories.

\subsec{Brane configuration} \subseclab{\braneconf}

\fig{A configuration which describes a six-dimensional gauge theory.
There are D6-branes which are represented by horizontal lines.
There are $n$ NS-branes represented by points.
The number of D6-branes in between two adjacent NS-branes is denoted by $V_i$.
D8-branes are represented by vertical lines. Their number is denoted by
$W_i$.}
{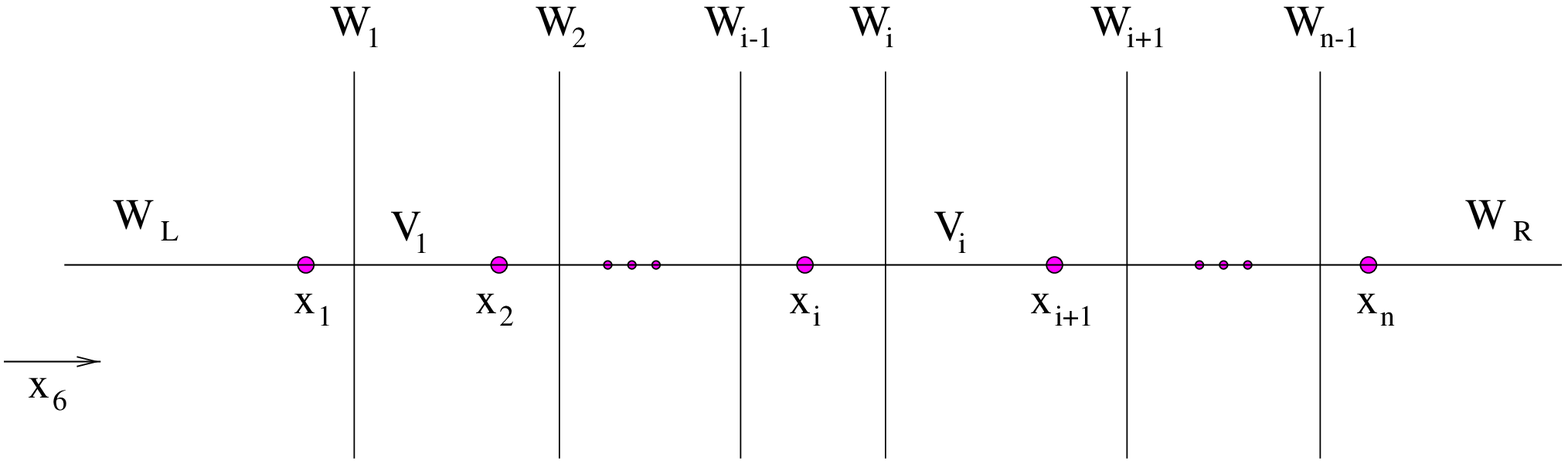}{16 truecm}
\figlabel\infin

Consider, as in figure \infin, a system with NS-branes with world-volume
$(012345)$,
D6-branes with world-volume $(0123456)$, D8-branes
with world-volume $(012345789)$. In some of the examples considered here, O8
orientifold planes parallel to the D8 and O6 orientifold planes parallel to
the
D6-branes will be introduced.

This configuration (in the case without orientifolds) can be considered as
the
T-dual of the system originally proposed in \hw\ and has been studied
in \refs{\karch,\hz}.

The presence of all these branes breaks spacetime Lorentz symmetry $SO(1,9)$
to
$SO(1,5)\times SO(3)$. In the directions $(x^0,x^1,x^2,x^3,x^4,x^5)$, which
are
common to all the branes,
a $(0,1)$ six-dimensional gauge theory is realized.
The $SO(3)$ symmetry acts on the 789 directions and is identified with the
$Sp(1)$ R-symmetry of the $(0,1)$ supersymmetry.
The D6-branes are stretched
in the direction $x^6$ and  the NS-branes are, in this direction, points
on
which the D6-branes are assumed to end.  The D8-branes are points in $x^6$
and
the 68-open strings give rise to hypermultiplets on the world-volume of the
D6-branes. A crucial difference with \hw\ is that the NS-branes are not
backgrounds for the gauge theory, but, filling exactly the directions
in which the six-dimensional theory is realized, provide tensor multiplets
to the theory. The D8 world-volume theory, having some infinite directions,
is still a frozen background for the
six-dimensional theory and appears as a global symmetry. In
\refs{\karch,\hz},
$x^6$ was taken to be non-compact to allow an arbitrary number of D8-branes. If
$x^6$
is compact, there are various cases to consider, which are restricted by charge
conservation. The familiar condition implies that we have exactly 16 dynamical
D8-branes and two orientifold planes, each carrying a charge $-8$ in units
where
the D8-brane carries charge $+1$.
In other words, we are dealing with a Type I$'$
background (on $S^1$).
A less familiar case is discussed in section 3.3.
In this section we consider a non-compact $x_6$
direction
in order to give the building blocks  for realizing different gauge
theories.
The compact case will be discussed in the next sections.
 
Let us consider, as in figure \infin, the case in which we have $N$ NS-branes
at
the positions $x_i^6$, $V_i$ D6-branes stretched between the $i$-th and the
$(i+1)$-th NS-brane and $W_i$ D8-branes in the $i-(i+1)$ segment.
We also put $W_L$ semi-infinite D6-branes on the left of the first NS-brane
and
$W_R$ semi-infinite D6 on the right of the last NS. The KK reduction of the
D6
world-volume theory along $x^6$ gives
rise to the gauge theory
\eqn\th{\prod_{i=1}^{N-1}U(V_i),}
with bifundamentals
charged under each two adjacent gauge group factors, $W_i$ fundamentals
for the $i$-th factor coming from the D8-branes and $W_{L,R}$ fundamentals
for the first and the last factor coming from the semi-infinite D6-branes.
Each of the NS-branes provides a $(0,2)$ tensor multiplet, but one of them
(the one that parametrizes the centre-of-mass motion of the system in the
directions transverse to the NS-branes) is decoupled.
There are 5 scalars in a $(0,2)$ tensor multiplet. The $N$ NS-branes give
rise
to $5N$ positions $x_i^\alpha$, $i=1,\ldots, N$, $\alpha=6,7,8,9,10$.
The scalars in the tensor multiplets $\phi_i^\alpha$ are proportional to
these positions in a relation that involves the string scale and the string
coupling
and will be written later.
Under $(0,1)$ supersymmetry each tensor multiplet decomposes into a $(0,1)$
tensor multiplet, which contains a real scalar $\phi_i^6$, and a
hypermultiplet,
which contains the rest of the 4 scalars $\phi_i^{7,8,9,10}.$
The last scalar of each hypermultiplet, $\phi_i^{10}$, lives on a circle of
radius $l_s^{-2}$.

The $x^6$ distances between the NS-branes are associated with the couplings of
the gauge factors by KK reduction \hw. The gauge coupling $g_i$ for the $i$-th
factor satisfies
\eqn\gaco{{1\over g_i^2}={x_{i+1}^6-x_i^6\over g_s l_s^3}=
\phi_{i+1}^6-\phi_i^6,}
where $g_s$ is the string coupling and $l_s$ is the string scale.
As noticed before, the position $x^6_i$ entering the gauge coupling is
related to
the scalar $\phi^6_i$ in the $(0,1)$ tensor multiplet.
Relation \gaco\ therefore provides  the standard coupling between $\phi$ and
the gauge fields,
\eqn\kar{\frac{1}{g^2} F^2_{\mu
\nu} + (\partial \phi)^2 + \sqrt{c} \phi F_{\mu \nu}^2,}
where $c$ is a numerical constant constrained by anomaly matching. Here
${1\over g^2}$ is a bare gauge coupling. With only one tensor
multiplet,
it can be reabsorbed in $\phi$.  

The 789 distances between the NS-branes are associated with the triplets of
FI
terms for the various $U(1)$ factors in the gauge group \hw. The $i$-th FI
term
$\vec w_i$ is given by
\eqn\fiterm{w_i^{1,2,3}={x_{i+1}^{7,8,9}-x_i^{7,8,9}\over l_s^3}.}

The last set of scalars gives rise to theta angles in six dimensions.
For each factor in the gauge group with a two-form field strength $F_i$,
if the quantity given by
$\tr(F_i\wedge F_i\wedge F_i)$ is
non-zero, there can be a term in the action with a coefficient
\eqn\thet{\theta_i={x_{i+1}^{10}-x_i^{10}\over l_s^3}.}

As a difference with \hw\ and the similar constructions in four and five
dimensions, the FI terms \fiterm\ and the theta angles \thet\ are dynamical
fields, being the four real scalar components of the hypermultiplets
provided by the NS-branes. They play a crucial role in cancelling the anomaly
of
the $U(1)$ factors \refs{\wpbl,\hm}. The $(0,1)$ supersymmetric completion of a
FI term gives rise to
a theta angle (eq. \thet ), which is shifted by a constant under a gauge
transformation. Its coupling to  $\tr(F_i\wedge F_i\wedge F_i)$ then
compensates
the $U(1)$ gauge anomalies. Moreover, the covariant kinetic term
\eqn\mass{\int d^6y (\partial_{\mu}\theta_i - A_{\mu i})^2}
makes the $U(1)$ gauge fields massive. In all the examples considered in this
paper there are as many FI hypermultiplets as $U(1)$ factors, and all the
Abelian anomalies are cancelled. As a consequence of this Green-Schwarz
mechanism, all the $U(1)$ factors are massive. All the $U(V)$ gauge factors in
the theories discussed in the following are therefore $SU(V)$ factors at
low
energy, even if we do not explicitly mention it.

In $(0,1)$ supersymmetric six-dimensional theories there are no BPS states
corresponding to particles. There are, however, BPS states corresponding to
strings. Their central charge formula is given by a linear combination of all
scalar fields in the various tensor multiplets.
In the brane configurations these BPS strings are realized by membranes
stretched along the $x^6$ direction in between two adjacent NS-branes. These
strings have a dual role. They are coupled to the tensor multiplets living on
the world-volume of the NS-branes and are thus charged both electrically and magnetically to the tensor multiplet. In addition, they are instantons to
the
gauge fields living on the D6-branes as being realized by D2-branes parallel to them.

We can obtain $SO$ and $USp$ gauge groups by introducing orientifold planes in
the picture. The simplest case is the introduction of an O6 plane. All the
branes must have an image under O6 or be stuck in 789. We will consider only the
case in which all the NS-branes are stuck. The generalization of \th\ is \jon\ :
\eqn\Oplane{...\times SO(V_{i-1})\times USp(V_i)\times SO(V_{i+1})\times
USp(V_{i+2})\times ...}
with bifundamentals
for each pair of neighbouring gauge group factors. There is a number of tensor
multiplets
equal to the number of gauge factors. The alternating presence of $SO$ and $USp$
is due to the fact that O6 changes its sign every time it crosses an NS-brane
\jon. There are no FI terms we can add from the field theory point of view, and
this indeed corresponds to the fact that the NS-branes are stuck in 789.

More interesting for us will be the introduction of O8 planes. Let us put an O8 plane at $x_6=0$. If $x_6$ is not compact there
is a possibility to choose the sign of the charge of the O8 plane \vafatre.
The
standard
charge ($-8$) is associated with an $SO$ gauge group for the D8-brane,
while
the choice of charge $+8$ (which we will refer to as O8$'$
in the following) 
corresponds to a $USp$ gauge group for the D8-branes. In the compact case,
there will be two cases to consider, which correspond to the introduction of
the
two types of orientifold planes restricted to the condition that the total
RR
charge must be zero. This will be discussed in the next section.

Each of the NS, D6 and  D8-branes now have  images under the $Z_2$ action
$x_6\rightarrow -x_6$. We now turn to identifying matter content corresponding
to different configurations.

\fig{$Sp$ ($SO$) gauge group coming from D6-branes stretched between NS and its
mirror under O8 (O8$'$) orientifold plane.}
{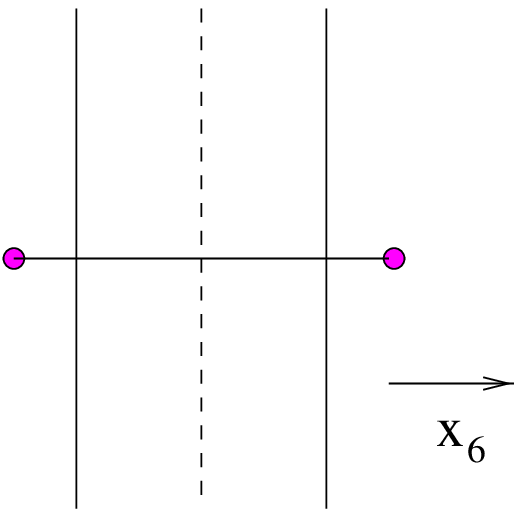}{6 truecm}
\figlabel\sp

\item{1.} The simplest configuration
corresponds to taking a single NS-brane located at a position away from the
$x^6$ origin and $V$ D6-branes stretched between it and its image.
The gauge group is $USp(V)$ ($SO(V)$ for O8$'$) and we can get $W$
fundamentals by putting $2W$ D8-branes ($W$ physical D8 and $W$ images).
The theory also contains the fields living on the NS world-volume, a $(0,1)$
tensor multiplet and a decoupled hypermultiplet.

\fig{$SU$ gauge group with a (anti)symmetric matter coming from D6-branes
stretched between the NS-brane stuck to the orientifold O8 (O8$'$) plane and another
NS-brane.}
{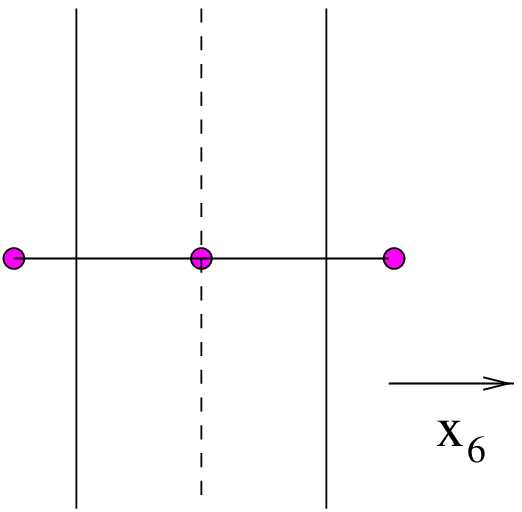}{6 truecm}
\figlabel\su

\item{2.} For infinite D6-branes this would be the end of the story, but,
since
our D6-branes are finite, we can construct a different configuration.
Let us stretch $V$ D6-branes between an NS-brane that is stuck at the
orientifold point and a dynamical NS-brane at the point $x^6_1$. The $Z_2$
projection maps the
D6-branes to their images (which connects the NS stuck at the fixed point
with
the image of the dynamical NS that lives at $-x^6_1$). In this way, the
orientifold projection identifies the fields on the finite D6-branes with the
fields on their images without further projecting the Chan-Paton factors and
we 
obtain a $U(V)$ gauge theory. The open strings, which connect the D6-branes and
their images, which carry Chan-Paton factors, become, after this
identification,
a hypermultiplet in the antisymmetric representation of $U(V)$ (or in the
symmetric representation for O8$'$)\LLnew.
The theory contains only one tensor multiplet. The $(0,1)$ tensor multiplet
associated with the stuck NS-brane in the $U(V)$ theory is projected out
since
its scalar $x^6$ is frozen. The 789 distance between the two NS-branes is an FI
term for the $U(1)$ factor in the gauge group, as in formula \fiterm. We can
move the stuck NS-brane in $789$ without breaking supersymmetry if, at
the same time, we
reconnect at the origin $x^6=0$ the D6-branes with their images, recovering
the
theory in item 1. From the field theory point of view, this corresponds to
giving an expectation value to the matter in the antisymmetric (symmetric) 
representation in order to compensate for the FI terms in the D-term equations, and this
indeed higgses $U(V)$ to $USp(V)$ ($SO(V)$).

We can generalize this construction by considering more NS-branes.
Consider first the case with $N$ NS-branes at the points $x^6_i>0$, with $V_i$
D6-branes and $W_i$ D8-branes between the $i$-th and $(i+1)$-th NS-brane. We
also put $W_R$ semi-infinite D6-branes on the right of the last NS-brane. There
are
moreover
$V_0$ D6 and $2W_0$ D8 (branes plus images) between the first NS in $x^6_1$
and
its image in $-x^6_1$. We have discussed before what happens to the $V_0$
D6-branes. For $i\ge 1$ the $V_i$ D6-branes are identified by the orientifold
projection with their images living in the negative $x^6$ axis, and there
is
no projection on the Chan-Paton factors. The field content is therefore
\eqn\theory{USp(V_0)\times U(V_1)\times U(V_2)\times ... \times
U(V_{N-1}),}
with bifundamentals for each pair of neighbouring gauge group factors, $W_i$
fundamentals
for the $i$-th factor and extra $W_R$ fundamentals for $U(V_{N-1})$. For
O8$'$ $USp$ is exchanged with $SO$. Consider next the case in which
there is an extran NS-brane stuck at the orientifold point, and
call $V_0$ and $W_0$ the number of D6 and D8 between this stuck NS and
the
first NS at $x^6_1$. The field content now is,
\eqn\theorytwo{U(V_0)\times U(V_1)\times U(V_2)\times ... \times
U(V_{N-1}),}
with bifundamentals for each pair of neighbouring gauge group factors, $W_i$
fundamentals
for the $i$-th factor, extra $W_R$ fundamentals for $U(V_{N-1})$  and an
antisymmetric (symmetric if we use O8$'$) for the
first factor $U(V_0)$.

We can generalize the previous theories by adding an O6 plane. In this case we
cannot have an NS-brane stuck at the O8 plane. In fact, the configuration with an
NS-brane and an O6 plane is not symmetric in $x^6$ around the location of the
NS-brane because the O6 sign is different on the left and on the right. In this
way, we have only one building block:

\item{1$'$.} For the case of a single NS-brane located at a position away from
the $x^6$ origin and $2V$ D6-branes stretched between it and its image, the
gauge group is $U(V)$ and we can get $W$
fundamentals by putting a total (physical plus images) of $2W$ D8-branes \gp.
Notice that the group of rank $V$ is obtained by using a total of $2V$ D6-branes. 
The theory contains also a $(0,1)$
tensor multiplet and a decoupled hypermultiplet.

The theories \theory\ are modified by the presence of the O6 plane in the following way:
\eqn\theoryO{U(V_0)\times USp(V_1)\times SO(V_2)\times USp(V_3)\times ...}
for the case of $N$ physical NS-branes. The last gauge factor
is $USp(V_{N-1})$ for $N$ even and $SO(V_{N-1})$ for $N$ odd. There are bifundamentals for each pair of neighbouring factors and $N$ tensor multiplets.
 $USp$ and $SO$ can be exchanged by simultaneously changing  the
overall sign for O6 and O8.

\subsec{Anomalies}\subseclab\anomalies

The matter content of an $N=1$ six-dimensional gauge theory is highly
constrained
by the gauge anomaly \refs{\seib,\dani} .\foot{All the non-decoupled $U(1)$
factors, which are  
automatically anomalous in six dimensions, are made massive by a Green-Schwarz
mechanism, as discussed in the previous section. In all the models we present
in
this paper, there are as many FI hypermultiplets (which can cancel the anomaly)
as $U(1)$ factors.} To obtain a sensible six-dimensional theory, the $F^4$ part
of the anomaly must identically cancel. If the remaining part of the anomaly
factorizes as
\eqn\an{\sum_{1=1}^P\left (\sum_a c_{ia}\tr F^2_a\right )^2}
with some real constants $c_{ia}$, it can be cancelled by $P$ tensor
multiplets
coupled as in \kar\ \sagnotti
\eqn\tens{\sum_{i=1}^P\sum_ac_{ia}\Phi_i\tr F^2_a .}

As a general rule, it is
believed that the charge conservation for the bulk fields corresponds to
the
anomaly cancellation on the world volume of the branes. We now derive the
charge conservation conditions for the bulk fields. As expected (and as can be
explicitly checked in all the examples presented in this paper), these
conditions
guarantee the cancellation of the quartic anomaly and the factorization of the
remaining part of the anomaly. Moreover, the brane system automatically
provides
the right number of tensor multiplets, coupled through \tens, to be able to
cancel  the anomaly completely.

The general charge
conservation conditions for the system presented in section \braneconf\ were
discussed in \hz. In general,
the charge of the finite D6-branes cannot flow inside the NS-brane on which
it ends (as happens in the T-dual model in \hw\ and in all its
generalizations
to other dimensions, from two to four) because the world-volume of the
latter
is too small. However, as discussed in \hz, the D8-branes induce a
cosmological
constant $m$ that modifies the equations of motion for the D6-branes with
terms containing the NS-NS antisymmetric tensor field. As a consequence,
an NS-brane can absorb a D6 charge equal to the value of the cosmological
constant at its position. In our normalization, the cosmological constant
is
an integer that jumps by one unit when a D8-brane is encountered. The general
condition is that at each NS-brane the difference between the number $n_L$
of
D6-branes ending on the left and the number $n_R$ of D6-branes ending on
the
right must be equal to the cosmological constant $m$:
\eqn\cons{n_L-n_R=m,}
$m$ is determined by the
number of D8-branes that are on the left of the NS-brane and the value of the
cosmological constant at $x^6=-\infty$. We will put the asymptotic value of the
cosmological constant equal to zero in the following.

Let us consider the simplest example, a $U(N_c)$
gauge theory with $N_f$ flavours. From the field theory point of view,
the
anomaly can be cancelled by adding a tensor multiplet if $N_f=2N_c$. This
gauge theory can be realized as in \th\ with $N=2$ by distributing the 
$N_f$
flavours between semi-infinite D6-branes and D8-branes. The two conditions
\cons\ read $W_L-V_1=0$, $V_1-W_R=W_1$ which indeed implies that
$N_f=W_L+W_1+W_R=2V_1=2N_c$. The tensor required to cancel the anomaly is
automatically provided by
the NS-branes. In the same way, $USp(N_c)$ and $SO(N_c)$ with $N_f$
flavours
can be obtained by \theory\ with $N=1$ and $N_f=W_1+W_R$. The value of the
cosmological constant at the position of the NS-brane is $W_1-8$ for O8 and
$W_1+8$ for O8$'$. The condition \cons\ correctly gives the anomaly
relation
$N_f=N_c+8$ for $USp(N_c)$ and  $N_f=N_c-8$ for $SO(N_c)$.

A second way of realizing $SO$ and $USp$ gauge theories is to use an O6
plane. Consider, for example, $N_c$ D6-branes between two NS-branes with
$W_{L,R}$ semi-infinite D6-branes in the
presence of an O6 plane. By choosing a negative sign for O6, the theory is
$SO(N_c)$ with $N_f=(W_L+W_R)/2$ flavours. The cosmological constant is zero and
therefore the D6 charge before and after each of the NS-branes must be the same.
The
O6 carries $-4$ units of charge\foot{We are using notations in which both a D6
and its image carry a unit of charge, since this convention simplifies the
charge conservation conditions. This must be contrasted with our notations for D8-branes, where a {\it physical} D8 carries one unit of charge.}. Since the
O6 sign is flipped by the NS-branes,  the charge of the O6 is $+4$ on the left of the first NS-brane and on the right of the
second. The charge conservation conditions are
therefore $+4+W_L=N_c-4=+4+W_R$, which once again provides the field theory relation $N_f=N_c-8$. 

In the general case \cons\ fixes the relation between number of colours and
number of flavours of the various gauge factors. The rules are simple. There is a relation for each
NS-brane which fixes the difference between the number of D6-branes on the left
and
on the right. The value of $m$
at that point is the total D8 charge existing on the left of that NS-brane,
computed with the rule that a D8 has charge $+1$, an O8 has charge $-8$ and
an O8$'$ has charge $+8$. If there is an O6 plane, it contributes to
the effective D6 number a factor $\pm 4$, where the sign is assigned in such a way that it is different for each adjacent gauge factor. We will soon see  the application of these
rules to more complicated examples. 
\subsec{Non-trivial fixed points}\subseclab{\nontriv}
The theories discussed in the previous sections have a Coulomb branch
parametrized by the real scalars in the tensor multiplets, or, in other
words,
by the relative $x^6$ positions of the NS-branes. At the origin of the Coulomb
branch, the NS-branes coincide in spacetime and we
expect tensionless strings. This is interpreted as the
signal for an interacting RG fixed point. The gauge factors associated with
pairs of coinciding NS-branes are automatically at strong coupling. For each
non anomalous theory discussed in this paper we can therefore expect the
existence
of a (in general) non-trivial fixed point at the origin of the Coulomb
branch.
The explicit construction of these theories, in terms of branes in a
consistent
string theory context, can be considered as a proof of the existence of such
fixed points.

Let us consider the fixed points in more detail.
In all the previous models there are BPS states associated to membranes
stretched in $x_6$ between the NS-branes. They appear in the six-dimensional
theory as strings with tension 
$1/g_i^2$, which can be naturally associated with instantons of the various
gauge factors. This interpretation is strengthened by the fact that a D2-brane
inside a D6-brane appears for the latter as a small instanton. At the origin
of
the
Coulomb branch, the positions of the NS-branes coincide and the strings
become
tensionless. The exotic six-dimensional theory living on the NS-branes reduces
to a standard non-Abelian gauge theory after compactification on a circle. $P$
coincident NS-branes
give rise to a five-dimensional $U(P)$ gauge theory. If $P$ NS-branes and their
images coincide at the O8 orientifold point, they give rise to a $USp(2P)$
gauge theory after
compactification\foot{The sign of the orientifold projection when acting on the
NS-branes, which determines the right symmetry between the two possibilities
$USp(2P)$ and $SO(2P)$, is actually difficult to establish in the perturbative
orientifold construction. We will see in section 5 that the system has an
interpretation as the world-volume theory of small $E_8\times E_8$ instantons.
In particular, in the strong coupling limit the system without D6-branes goes
smoothly to
the theory of $P$ $E_8\times E_8$ instantons in flat space, which is known to be
a theory of P tensor multiplets with a symmetry $USp(2P)$. The presence of D6-branes cannot affect the way in which the orientifold projection acts on the NS-branes. A different argument, based on the structure of the Coxeter box, is
given in section 3.3.}.

The theories \th, \theory, \theorytwo\ (provided that, in all cases, the $V$'s and
$W$'s are chosen
in such a way as to cancel the anomaly\foot{ The precise relation is discussed in
sections 3 and 4.}) give rise to strong coupling fixed points with the same
gauge and matter content and with $N$ tensor multiplets associated with
a symmetry $U(N)$, $USp(2N)$ and $USp(2N-2)$ \foot{We are assuming that the
presence of stuck NS-branes does not change the symmetry. This assumption
agrees with results in \intr\ and observations in section 3.3, but, at the
moment, we have no arguments for ruling out other possibilities.}, respectively.  In the case of
\th\ 
one of the tensor multiplets is decoupled and the
symmetry is actually $SU(N)$. In the same way, the theories \Oplane, \theoryO\
give rise to strong coupling fixed points with the same
gauge and matter content and with $N$ tensor multiplets associated with
a symmetry $U(N)$, $USp(2N)$, respectively. 

Some of these fixed points were found in \ken. The precise relation
between the two approaches is discussed in section 4. The brane construction
makes
manifest the tensionless string symmetry, which was discussed in \intr.

\newsec{The compact case}
\subsec{The Type I$'$ background}\subseclab{\typeIp}

Let us go back to the case with compact $x^6$. The total D8 charge in the
compact
direction must sum up to zero. There are two cases to consider. The more
familiar case is the introduction of orientifold planes of
negative charge. We can consider this background as Type IIA/$\Omega Z_2$,
where
$\Omega$ is the world-sheet parity and $Z_2$ acts as $x^6\rightarrow -x^6$.
This background is known as Type I$'$ on $S^1$, the T-dual of the Type I
theory.
We have two orientifolds O8
at $x^6=0$ and $x^6=\pi R$. We therefore need 16 physical D8-branes to
compensate the charge. In the picture in which $x^6$ is a circle with
opposite
points identified by $Z_2$, for each
physical D8-brane there is an image under $Z_2$. When all D8-branes are on top
of one of the O8 planes, they
give rise to the space-time gauge group $SO(32)$.
In the second case we put an O8 at $x^6=0$ and an O8$'$ at $x^6=\pi R$.
No additional branes are needed to cancel the O8 charge. This case is discussed
in section 3.3.

Consider the theory obtained by putting $N$ NS-branes on the circle and D6
branes stretched between them. The NS-brane positions must respect the $Z_2$
symmetry. If $N=2P$ we can put $P$ physical NS-branes at arbitrary points
and
other $P$ NS-branes at the image points. The world-volume fields are
correspondingly identified and the $N$ NS-branes provide only $P$ tensor
multiplets to the six-dimensional theory.
If $N=2P+1$, one of the NS-branes must
be stuck to one of the orientifold points. Since the frozen position $x^6$
is
the scalar partner of the tensor field on the world-volume of an NS-brane,
the
tensor multiplet of the stuck NS-brane is projected out and the total number
of tensor multiplets is still $P$. The gauge content of these models can be
easily determined with the rules in section \braneconf\ and reads
\eqn\vec{\eqalign{USp(V_0)&\times U(V_1)\times U(V_2)\times\cdots
U(V_{P-1})
\times USp(V_P),\,\,\, N=2P,\cr USp(V_0)&\times U(V_1)\times
U(V_2)\times\cdots
U(V_{P-1})\times U(V_P),\,\,\, N=2P+1,}}
with bifundamentals for each pair of neighbouring gauge factors, an
antisymmetric for the last $U(V_P)$ factor for $N=2P+1$ and $W_i$ 
fundamentals for the $i$-th factor coming from the 6-8 strings. 
The theories are coupled to $P$ tensor multiplets. Each of the NS-branes
provides also a hypermultiplet, which is parametrized by the position of the
brane in $(7,8,9)$ and in the 10th M-theory direction. The positions in
$(7,8,9)$ of the NS-branes are, as in \fiterm, FI terms for the gauge theory.
The 10th positions are theta angles as in \thet.
We see that we have exactly the same number of FI hypermultiplets surviving
the
$Z_2$ projection\foot{By translational invariance in the 789 and 10 directions,
one of the hypermultiplets
coming from the NS-branes is always decoupled.} and of $U(1)$ factors in the
gauge theory ($P-1$ for $N=2P$ and $P$ for $N=2P+1$). The FI hypermultiplets
play the crucial role of rendering massive the $U(1)$ factors that in six
dimensions
are necessarily anomalous, as discussed in section \braneconf.

The global symmetry coming from the D8-branes in these models is
\eqn\glob{\eqalign{SO(W_0)&\times U(W_1)\times\cdots U(W_{P-1})
\times SO(W_P),\,\,\, N=2P,\cr SO(W_0)&\times U(W_1)\times\cdots
\times U(W_P),\,\,\, N=2P+1.}}
There is a third non-Abelian symmetry for this set of models (the other two are
the gauge group and global symmetry coming from D6- and D8-branes,
respectively.)
This symmetry is associated with coinciding NS-branes. In this
configuration, the light states are strings, which transform in the adjoint representation of the
corresponding group. They appear at the fixed points of these theories and will
be discussed in section 3.3.

In the case in which $N$ is even, there is a second configuration that is invariant
under $Z_2$. We can put one stuck NS-brane at each of the orientifold
planes.
This results in the following theory
\eqn\wvec{U(V_0)\times U(V_1)\times ... U(V_{P-2})\times U(V_{P-1}),\,\,\,
N=2P,}
with bifundamentals for each pair of neighbouring gauge factors, an
antisymmetric for the first and last $U$ factor, and $W_i$ 
fundamentals for the $i$-th factor. The theory is coupled to $P-1$ tensors,
one
for each of the dynamical NS-branes. The number of $U(1)$ factors is $P$, which
is the same as the number of FI hypermultiplets provided by the NS-branes.

In all these theories, the total number of
fundamentals $\sum W_i$ is equal to 16, the number of physical  D8-branes.
$W_i$ and $V_i$ must satisfy a particular relation in order to cancel the gauge
anomalies. The relation can easily be determined using the rules in section
\anomalies. Let us determine, for example, the anomaly constraints for
$N=3$.
The theory is $USp(V_0)\times U(V_1)$ with $W_0$ fundamentals for
$USp(V_0)$,
$W_1$ fundamentals and an antisymmetric for $U(V_1)$. It is realized by
stretching
$V_0$ D6-branes between the first orientifold and a dynamical NS-brane, and
$V_1$ D6 between the dynamical NS and a second NS-brane stuck at the
second
orientifold plane. $W_0$ D8-branes live before the first NS and $W_1$
after,
with the constraint that $W_0+W_1=16$. The anomaly constraints are derived by
imposing the relation \cons\ at the position of the dynamical NS-brane. The
cosmological constant at that point is easily computed to be $-8+W_0$ (the
total D8 RR charge on the left of the NS-brane: $-8$ is the contribution of the
leftmost O8 and $W_0$ is the contribution of the D8-branes). The constraint is
therefore
\eqn\constrain{V_0-V_1=-8+W_0,}
and the final result for the gauge theory is $USp(2k)\times U(2k+W_1-8)$,
where we redefined $V_0=2k$.

The general result for \vec\ and \wvec\ can easily be derived and reads,
\eqn\ano{V_i=2k + \sum_{j=1}^{j=P}\min(i,j)W_j-8i,\,\, i=0,...,P,}
for \vec, and
\eqn\anow{V_i=2k + \sum_{j=1}^{j=P-1}\min(i,j)W_j-8i,\,\, i=0,...,P-1,}
for \wvec.

That these relations ensure the complete anomaly cancellation for the
theories
\vec\ and \wvec\ was explicitly checked in \ken, where these theories were
first introduced as the world-volume theories for small $SO(32)$ instantons
living on top of orbifold singularities. We discuss in detail the relation
of our system with small $SO(32)$ instantons in section 4.
Note that for the even case of eq. \vec, the last group factor requires
that $V_P$ is an even number. This is because the orientifold projection
requires an even number of
D6-branes if they are the first or the last factor on the segment.
Note that $V_0$ is chosen in such a way.
This condition imposes a restriction on the possible distribution of the
D8-branes on the interval. Using eq. \ano\ we find
\eqn\anoa{V_P=2k+\sum_{j=1}^PjW_j-8P.}
This implies that the condition for $V_P$ to be even is that
\eqn\anob{\sum_{2j+1\le P}W_{2j+1}}
is even.

Let us study some dynamics for these gauge theories.
We will start with the configuration with no NS-branes and proceed by adding
them. In field theory this corresponds to including more and more matter
and, as we will find, an inverse Higgs mechanism. This is done by moving an
NS-brane in the 789 directions. This transition is allowed only if the resulting
configuration is consistent with charge conservation. We should also note that
an NS-brane moving in a background with non-zero cosmological constant of value
$m$ is bound on the left to $m$ D6-branes. So when moving an NS-brane with $m$ D6-branes attached, we are changing the gauge group content.

The configuration of $k$ D6-branes, stretched between the two O8 planes, with
their images gives rise to a $USp(k)$ gauge group with matter in the antisymmetric
representation of the gauge group and 16 flavours in the fundamental
representation. There is one background (non-dynamical in the six-dimensional sense) tensor multiplet corresponding
to the length of the interval that serves as a gauge coupling.
The antisymmetric representation corresponds to the motion of the D6-branes in the
789 directions.
There are few cases to consider when the NS-brane is moved from infinity in the
789 and 10 directions.
1) The NS-brane is stuck at one of the orientifolds.
In this case the resulting gauge
theory is $U(2k)$ with matter in the antisymmetric and 16 fundamentals.
For the $U(2k)$ theory the process of moving away the NS-brane corresponds to
turning on an FI term, which induces an expectation value for the antisymmetric
breaking down to the $USp(2k)$ gauge group.
2) The NS-brane is away from the orientifold, moving along with its image.
Suppose that the NS-brane has $W_0$ D8-branes to its left and $W_1$ D8-branes to
its right. The cosmological constant felt by the NS-brane is $-8+W_0$.
Correspondingly, there must be $-8+W_1$ long D6-branes ending on the NS
brane from the right. The resulting gauge group is $USp(2k)\times USp(2k-8+W_1)$. More
generally, we could have moved the NS-branes with $2r$ D6-branes on its left
and $2r-8+W_1$ on its right, with a final gauge group
$USp(2k+2r)\times USp(2k+2r-8+W_1)$. This time, the process of moving the
NS-brane cannot correspond to turning
on an FI, since there are no $U(1)$ factors in the gauge group, but it is still an inverse Higgs mechanism. The theory $USp(2k+2r)\times USp(2k+2r-8+W_1)$  indeed has flat directions where the matter field in the
bifundamental gets an expectation value, breaking the group to $USp(2r)\times USp(2r-8+W_1)$ with the usual matter content and $USp(2k)$ with an antisymmetric and 16 fundamentals, which was diagonally embedded in the two original groups. This is exactly the process we just described in terms of branes. Anticipating the results of the next section, we can interpret this transition as
follows. The theory $USp(2k+2r)\times USp(2k+2r-8+W_1)$ describes $k+r$ instantons sitting on a $Z_2$ singularity. The Higgs transition corresponds to moving
$k$ small instantons away from the singular point. $USp(2k)$ with an antisymmetric and 16 fundamentals is just the theory of $k$ small instantons in flat space.

Moving in or away an NS-brane with the right number of attached D6-branes so as to
guarantee charge conservation, gives the result expected from the field
theory point of view. It is, in general, a Higgs mechanism. If the NS-brane
position is coupled to the theory as in \fiterm, the Higgs mechanism is induced
by the turning on of an FI term. In the remaining cases, it is associated with 
flat directions. By moving NS-branes we can connect all the theories in \vec\ and \wvec.

Consider, for example, the $N=5$ case in \vec. This corresponds to having one NS-brane at one orientifold and 
other two NS in the middle of the interval. The group is $USp(2k)\times U(2k+W_1+W_2-8)\times U(2k+W_1+2W_2-16)$. Assume, for simplicity, $W_2>8$. The other case can be discussed in a similar way. If we move away  the
middle NS-brane in the 789 directions (this corresponds to turning on an FI term), $W_2-8$ D6-branes must be attached to it on the right to compensate the
cosmological constant at that point. The resulting theory is factorized in
$USp(2k)\times U(2k+W-8)$
(where $W=W_1+W_2$ is the total number of D8-branes between the two remaining NS
branes) and $USp(W_2-8)$. We have obtained the cases $N=3$ and $N=2$ in \vec. 
The two theories are decoupled for sufficiently large distances. In the language of the next section, we started with $k$ instantons with vector structure sitting on a $Z_5$ singularity, and we turned on some blowing up parameter. The
singularity is deformed to a pair of (far enough) $Z_3$ and $Z_2$ singularities. The $k$ instantons are sitting on the $Z_3$ singularity. Some other finite D6-branes are stuck on the $Z_2$ singularity, and cannot leave the singular point. More generally, we could have moved the NS-brane with $k_2$ D6-branes on its left and $W_2+k_2-8$ on its right. This corresponds to a different solution of
the D terms equation of the gauge theory in the presence of an FI term. The resulting theory is factorized in $USp(2k_1)\times U(2k_1+W-8)$ and $USp(2k_2)\times Usp(2k_2+W_2-8)$ ($k_1+k_2=k$) and can be interpreted as a splitting of the $Z_5$ singularity in $Z_3\times Z_2$ while keeping $k_1$ instantons on the $Z_3$ singularity 
and $k_2$ on the $Z_2$.

There is a second possibility to consider in the same example. We can move the stuck NS-brane (turning on an FI term for the
last $U(1)$ factor). We do not need to attach D6-branes to the NS because no
charge conservation condition is associated to the NS-branes stuck at the
orientifold points. As already discussed in section \braneconf, the only effect
of moving the stuck NS-brane is to higgs the corresponding $U(V)$ factor
to $USp(V)$. In this way, the case $N=5$ is higgsed to the case $N=4$ in \vec.
We see that all the theories in \vec\ can be connected by turning on or off
some FI terms. More generally, even the theories in \wvec\ can be connected to
those in \vec\ by turning on FI terms. Consider, in fact, the theory with $N=4$
in \wvec. There are three NS-branes, and two of them are stuck. If we move the
middle one, the theory is higgsed to the case $N=2$ in \wvec. If we move one
of the stuck NS, however, the theory is first higgsed to the $N=3$ case in
\vec, which can be successively connected to the case $N=2$ in \vec. We see, in
conclusion, that all the theories in \vec\ and in \wvec\ can be connected by
turning on or off some FI terms. The general picture should be clear from the
previous examples. The process of moving NS-branes corresponds to a Higgs mechanism for which some instantons can move away from the singularity (finite D6-branes stretched  between NS-branes
can reconnect to infinite ones, which can live without ending on the NS, and
therefore far from the singularity) or to the turning on of some FI term, which splits the singularity (the instantons can be partitioned among the singularities or moved far from all of them).

\subsec{Small Instanton Transition}\subseclab{\trans}
A transition that describes a trade of a tensor multiplet for 29
hypermultiplets is familiar in the physics of small $E_8$ instantons. When the instanton becomes of zero size, 29 hypermultiplets are eaten by
a tensor multiplet, in a sort of inverse Higgs mechanism. As we already anticipated, the theories we are considering describe small $SO(32)$ instantons sitting at spacetime singularities, and the presence of tensor multiplets suggests
that the same phenomenon also happens in this case. We have no Lagrangian description for such a transition. The conclusion that
a tensor is traded for exactly 29 hypermultiplets usually follows from the fact that both make the same contribution to the gravitational anomaly. 
We now want  to show that this transition can be explicitly demonstrated in
the brane system.

Consider the case $N=2$ in eq. \vec.
Let us put all the 16 D8-branes to the left of the NS-brane.
The gauge group is $USp(2k)\times USp(2k-8)$ with 16 hypers in the fundamental
of the first group and a bifundamental.
The NS-brane is not allowed to move in the 789 and 10 directions, as this would
violate charge conservation.
We can tune the position of the NS-brane to move towards the right orientifold
plane.  For simplicity, let us choose $k=4$.
For this case there are no D6-branes to the
right of the NS-brane.
At the fixed point, the 4 D6-branes can move freely along the 789
directions and thus provide more moduli. There are 27 + 1 hypermultiplets
corresponding to the antisymmetric representation of $Sp(4)$. In addition one
NS-brane can move freely in the 789 and 10 directions, providing a further
hypermultiplet. At this point the other NS-brane is not allowed to move in the
$x^6$ direction and the tensor multiplet is thus frozen. We have just described, as promised, a transition in which a tensor multiplet is traded for 29 
hypermultiplets. The case $k>4$ gives rise to a transition to the theory
$USp(2k)$ with an antisymmetric and 16 fundamentals. Once again this theory
has 29 extra hypermultiplets, one of which is provided by the relative motion
of the NS-branes in the 789 and 10 directions.
It is now easy to generalize this transition to general $N$. We can tune the
rightmost NS-brane to move to the right orientifold plane. At the point where
the NS-brane is approaching its image with respect to the orientifold plane, there
are tensionless strings indicating a non-trivial fixed point. The NS-branes can
separate at the orientifold plane and are no longer allowed to move in the
$x^6$ direction. A tensor multiplet is frozen. The number of gauge factors is
reduced by one, and the last gauge group gets additional matter corresponding to
the new configuration. A small instanton transition has been performed.

\lref\aspin{P.S. Aspinwall, {\it Point-like Instantons and the Spin(32)/Z2 heterotic String}, Nucl. Phys. B496 (1997) 149, hep-th/9612108.}

The specific example
with $Sp(4)$ 
was first discussed in \aspin\ and further in \ken.

\subsec{Generalizations}\subseclab{\generalisation}

Let us turn to the second case, in which there is an O8 at $x^6=0$ and an
O8$'$
at $x^6=\pi R$. There are no additional D8-branes since the charge sums to
zero.

Consider, as before, $N$ NS-branes on the circle and D6-branes stretched
between
them. The gauge group is now
\eqn\veci{\eqalign{USp(V_0)&\times U(V_1)\times U(V_2)\times\cdots U(V_{P-1})
\times SO(V_P),\,\,\, N=2P,\cr USp(V_0)&\times U(V_1)\times
U(V_2)\times\cdots
U(V_{P-1})\times U(V_P),\,\,\, N=2P+1.}}
with bifundamentals for each pair of neighbouring gauge factors, and
symmetric for the last $U(V_P)$ factor for $N=2P+1$. 
The theories are coupled to $P$ tensor multiplets and  we have exactly the
same
number of FI hypermultiplets surviving the $Z_2$
projection as $U(1)$ factors in the gauge
theory ($P-1$ for $N=2P$ and $P$ for $N=2P+1$).

In the case in which $N$ is even, there is a second configuration that is invariant
under $Z_2$. We can put one stuck NS-brane at each of the orientifold
planes.
This results in the following theory
\eqn\wveci{U(V_0)\times U(V_2)\times ... U(V_{P-2})\times U(V_{P-1}),\,\,\,
N=2P,}
with bifundamentals for each pair of neighbouring gauge factors, an
antisymmetric for the first and a symmetric for the last $U$ factor.
The theory is coupled to $P-1$ tensors, one for each of the dynamical NS-branes.
The number of $U(1)$ factors is $P$, which is
the same as the number of FI hypermultiplets provided by the NS-branes.

Formulas \ano\ and \anow\ apply for this case by setting all $W$'s to zero:
\eqn\anoo{V_i=2k -8i,\,\, i=0,...,P\,\,\, (P-1).}
Arguments similar to those given in the previous section show that all the
theories in \veci\ and in \wveci\ can be connected by turning on or off FI
terms.
\lref\kuta{D. Kutasov, {\it Orbifolds and Solitons}, hep-th/9512145.}
\lref\sen{A. Sen, {\it Duality and Orbifolds}, hep-th/9604070.}

A different generalization involves introducing an extra O6 plane in the game.
Since we cannot have stuck NS-branes, as discussed in section \braneconf\ ,
we have only the case $N=2P$. The theory is
\eqn\theoryOO{U(V_0)\times USp(V_1)\times SO(V_2)\times USp(V_3)\times ...\times U(V_P),}
where the gauge factor before $U(V_P)$
is $USp(V_{P-1})$ for $P$ even and $SO(V_{P-1})$ for $P$ odd. There are bifundamentals for each pair of neighbouring factors and $P$ tensor multiplets.
 $USp$ and $SO$ can be exchanged by simultaneously changing the
overall sign for O6 and O8.   
 The anomaly-free value for the $V$'s can be determined in the usual way
by the charge conservation at each NS-brane, adding the only new ingredient that
the O6 plane contributes $\pm 4$ to the D6 charge and that the O6 sign flips
every time one crosses an NS-brane. It should also be  remembered that a
factor $U(V)$ coming at one O8 plane in the presence of an O6, by construction,
is obtained with $2V$ D6-branes.

\subsec{Fixed points, Coxeter boxes and string theories}\subseclab{\fix}

All the theories discussed here describe at the origin of the
Coulomb branch a non-trivial RG fixed point, associated with tensionless
strings.
The origin of the Coulomb branch corresponds to putting all the dynamical
NS-branes on one of the orientifold points, for example the left one. All the
gauge factors, except the
last associated to the other orientifold, are at strong coupling. The last
gauge factor has instead a gauge coupling proportional to the radius of the
$x^6$ direction and, being IR-free, becomes ungauged at the fixed point.
A similar argument applies if the NS-branes coincide at the other orientifold.
In this way, the first or the last factor must be dropped in the description
of the gauge and matter content at the fixed point. The possible non-trivial
fixed points were therefore already described in section \nontriv.

We should note that, when $x_6$ is compact, the scalars in the
tensor
multiplets are also compact. In general, they must parametrize the Coxeter box
of the simple group associated with the tensors via the relations \intr
\eqn\cox{\alpha_{\mu}\times \phi + L\delta_{\mu 0}\ge 0,\,\,\, \mu =0,...,r}
where $\alpha_{\mu}$ are the simple roots of the group, $r$ its rank and the
extended root $\alpha_{0}$ is defined by $\sum_0^r n_{\mu}\alpha_{\mu}=0$
($n_\mu$ are the Dynkin indices), while $L$ is the size of the Coxeter box. 

For the theories discussed in the previous sections, the size of the Coxeter box is
determined by the radius of the compact $x_6$ direction. Equation \gaco\ gives 
\eqn\lengcox{L={R_6\over l_s^3g_s}.}

The natural disposition of $r$ NS-branes on the segment ($r=P$ for \vec, \veci\ and $P-1$ for \wvec, \wveci, for example) gives the relations
\eqn\coxtwo{0\le \phi_1,\qquad \phi_i\le\phi_{i+1},\qquad \phi_r\le L,}
which exactly corresponds to eq. \cox\ when the group $USp(2r)$ is used.
The use of $SO(2r)$ in \cox\ would give a set of relations that are not compatible with \coxtwo. We can consider this as  further evidence for the fact that
the NS-branes living on the segment parametrize the Coxeter box of $USp(2r)$.

To provide a full definition for the theories discussed in this section, we need
to find a six-dimensional string theory, with gravity decoupled, which
contains  our theories at low energy. We postpone the discussion of how to 
find such a string theory until section 6, since we need first to understand the effects of
a string duality on our system.

\newsec{$SO(32)$ instantons in Type I}

The theories \vec, \wvec\ were first introduced in \ken\
studying small $SO(32)$ instantons living on top of
orbifold singularities. Are there some small instantons in our
construction of six-dimensional theories? Small instantons are represented
in
Type I by D5-branes. If we make a T-duality in $x^6$, the Type I$'$
background
becomes the Type I theory, our D6-branes are turned
into D5-branes, and the $N$ NS-branes become a spacetime $Z_N$ orbifold
singularity \vo. Our model, in the T-dual picture, is equivalent to the
theory
of small $SO(32)$ instantons living on top of an orbifold singularity.

Let us see how the same theories are realized in Type I \ken. Consider $k$ D5-branes
living on top of a $Z_N$ singularity.
A small instanton is specified
by the instanton number $k$ and by the value of a non-trivial flat
connection
at infinity, which is a representation of the orbifold singularity $Z_N$ in the
gauge group. Since the string theory gauge group is spin$(32)/Z_2$, the
gauge group element $\rho_{\infty}$, which specifies the flat connection at
infinity, can satisfy both $\rho_{\infty}^2=-1$ (which we will refer to as
the
case without vector structure) or $\rho_{\infty}^2=1$ (case with vector
structure) \wpbl. The difference between the two cases is implemented in
the
perturbative
orientifold construction as a different action for the world-sheet parity
$\Omega$ on the Chan-Paton factors \kenju\foot{This results in different
commutation relations between
$\Omega$ and the generators of the 
orbifold singularities.}. The possible world-volume gauge theories living on
the
D5-branes were
classified in \hm\ for a $Z_N$ singularity and generalized to other kinds of
singularities in \refs{\ken,\kenju,\kenjutwo}. The final result is the
theory
\vec\ for the case with vector structure and the theory \wvec\ for the
case
without vector structure (for even $N$). We see that the two cases
 were described in the Type I$'$ picture in a unified way. 

The fields living on the NS-branes become, after T-duality, the twisted fields
coming from the $Z_N$
orbifold projection. Each of the twisted sectors usually provides, before
projecting with $\Omega$, a $(0,1)$ tensor multiplet and a hypermultiplet,
which is
exactly the world-volume content of an NS-brane. In Type I$'$,
$\Omega Z_2$ projects out some of the tensors and some of the
hypermultiplets.
In  Type I, on the other hand,
we expect that all the tensors are projected out by $\Omega$ 
and that all the hypermultiplets survive and parametrize the blowing up modes that
can smooth the
singularity.  Such a naive Type I picture could not provide the tensor multiplets
existing in
the theories \vec\ and \wvec. It turns out, however, that with the standard
perturbative definition
of $\Omega$, the orientifold is not the same as  the Type I theory compactified
on the corresponding non-singular manifold.  It was realized in
\refs{\sagtwo,\sagb,\sagbtwo,\gj,\pol,\kenju} that the perturbative definition of $\Omega$ projects
different twisted sectors 
in a different way and some tensor multiplets survive. These twisted fields
live in six dimensions and must be included in the six-dimensional theory,
providing the right number of tensor multiplets in \vec\ and \wvec.    
The surviving hypermultiplets  appear on the D5 world-volume theory as FI terms
and take care of the anomalous $U(1)$ factors \refs{\wpbl,\hm}. The final
result agrees with what we found in the T-dual picture.  

The total number of
fundamentals $\sum W_i$ is equal to 16, the number of physical (without images
under $\Omega$) D9-branes, and the way in which 16 is partitioned among the
$W_i$ specifies the way in which the flat connection at infinity breaks
$SO(32)$.

The number of FI terms we can add in this theory is not the total number
$N-1$
of blowing up modes of the singularity. This means that some blowing up
modes are frozen and the manifold cannot be made completely smooth. The
missing
modes are traded for tensor multiplets.
It is conjectured in \ken\ that these theories flow to non-trivial interacting
fixed points and that, at the origin of their Coulomb branch, there is a
transition in
which each tensor is traded for 29 hypermultiplets. We gave an explicit brane
description of such a transition in section \trans. After the transition, the
total number of hypermultiplets in the Higgs phase is equal to the
dimension
of the relevant moduli space of $k$ $SO(32)$ instantons on a $Z_N$ singularity.

All the small instantons theories, with or without vector structure, can be
connected by turning on or off some FI terms, or, in geometrical words, by
turning on or off some blowing up parameters. This was perhaps clearer in the
T-dual picture of section \typeIp, where the theories with and without vector
structure appeared in a unified description. Translating the results of section
\typeIp\ in the language of the small instantons, we see that there can be
transitions between instantons without vector structure and instantons with
vector structure when we turn on some blowing up parameters and smooths down the
singularity.
As we saw in the previous section, for example, the case of instantons without
vector structure at a
$Z_4$ singularity reduces, according to which blowing up parameters we turn on,
both to the case of instantons without or with vector structure at a $Z_2$
singularity\foot{For arriving at the $Z_2$ case without vector structure, we
must first go through the case of a $Z_3$ singularity.}.

The relations \ano\ and \anow\ were explicitly checked, from the field
theory
point of view, in \ken,  and derived using 
the space-time charge conservation in \kenju. The D5-branes are charged under
the six-dimensional twisted fields and an explicit
calculation of the twisted tadpoles can reproduce \ano\ and \anow. The
calculation involves the evaluation of partition functions and it is
considerably longer than the charge conservation argument we used in the
T-dual
picture.

\newsec{$E_8\times E_8$ instantons}

Useful information  on the gauge theories defined as configurations of branes
were often obtained by exploring the strong coupling limit of the string
theory
in which they are embedded \refs{\hw,\witten}. The strong coupling limit of
our
Type I$'$ background is well described, for a generic configuration of D8-branes, by
a weakly coupled $SO(32)$ heterotic theory. In this limit, D-branes are no
longer
a tractable object and we cannot hope to gain new information. However, in
nine dimensions, the heterotic $E_8\times E_8$, which admits a completely
different strong coupling limit as M-theory on a segment, lives in the same
moduli space. Type IIA on $S^1/\Omega Z_2$ can indeed be considered as
M-theory on $S^1/Z_2\times S^1$, which is the strong coupling limit of the
heterotic $E_8\times E_8$ compactified on $S^1$. In this identification,
$x^6$
is exchanged with $x^{10}$ in representing the eleventh dimension, which opens
up
in the strong coupling limit. 
In this M-theory picture, NS-branes become M-theory five-branes and $K$
parallel
infinite D6-branes become a KK monopole. The KK monopole determines a non-trivial Taub-Nut metric in $(7,8,9,10)$, which reduces
in the infinite string coupling limit to an ALE singularity of type
$Z_K$.
The five-branes live at a point in $x^6$ (which is the segment on which M-theory
is defined) and at a point in the Taub-Nut space. They could be naively
interpreted as small $E_8\times E_8$ instantons that have left the
boundary.
The fate of a configuration of D6-branes stretched between NS
branes
is a more delicate question.

To see if the interpretation of the system as describing small $E_8\times E_8$
instantons makes any sense, let us analyse the case in which the
transition from Type I$'$ to M-theory can be completely understood \branes .

The two heterotic theories are T-dual in nine dimensions. Consider a subgroup
$SO(14)$ of $SO(32)$. At a special point in moduli space, the $SO(14)$ and a
$U(1)$ coming from the KK reduction are enhanced to $E_8$ by massless winding
modes of the heterotic string. Under heterotic/Type I$'$ duality, the same
point
in moduli space corresponds to having
seven D8-branes at each of the orientifold points and the remaining two in a
symmetric way on the segment. Upon duality, the heterotic winding modes are
mapped to zero branes stuck at the orientifold points. At the special point
in
which we expect enhanced symmetry, the Type I$'$ coupling constant diverges
at
the orientifolds, and the zero-branes become massless providing the missing
gauge bosons for realizing $E_8\times E_8$. 

Consider therefore the configuration with seven D8-branes at each of the
orientifold
points and put $P$ physical NS-branes on the segment.
Between the two D8-branes that are not at the orientifold, the cosmological constant is
zero. If we have several NS-branes in this region, we can stretch the same
number ($K$) of D6-branes between them. On the left of the first D8-brane, the
cosmological constant is $-1$; going towards the left, the number of D6-branes should decrease
by one unit from gauge factor to gauge factor. The same
happens on the right of the second D8, where the cosmological constant is
$+1$. If $P\ge 2K$ we eventually end with a $U(1)$ theory on the extreme
left
and right, where the first and last NS-branes live. The gauge content is
therefore
\eqn\asp{U(1)\times U(2)\times ...\times U(K-1)\times U(K)^{(P-2K+1)}\times
U(K-1)\times ... \times U(1),}
with bifundamentals for each pair of each neighbouring gauge factors
and
one fundamental for the first and last $SU(K)$ factor coming from the two D8
branes. There are $P$ tensor multiplets. Taking into account that the $U(1)$
factors are massive for a Green-Schwarz mechanism \wpbl, we can rewrite the
previous theory as

\eqn\asptwo{SU(2)\times SU(3)\times ...\times SU(K-1)\times
SU(K)^{(P-2K+1)}\times SU(K-1)\times ... \times SU(2),}
with bifundamentals for each pair of neighbouring gauge factors and
one fundamental for the first and last $U(K)$ factor coming from the two D8-branes and one fundamental for the two $SU(2)$ coming from the two frozen
$U(1)$
factors.

This theory was indeed proposed in \am\ as the theory of $P$ 
$E_8\times E_8$ small instantons living at a singularity of type $Z_K$. We can
explicitly see how this configuration is lifted to the standard M-theory
picture of $E_8\times E_8$ small instantons. In the region between the two
D8-branes, the background theory is of Type IIA. An eleventh dimension is
supposed
to decompactify in the strong string coupling limit, and the two D8-branes move
toward an end \branes, creating the two M-theory boundaries described in
\horava.
Meanwhile, the D8-branes have to cross a certain number of NS-branes.
As described in \hz, every time a D8-brane crosses an NS-brane a new D6 is
created between them. By the time the D8-branes touch the orientifolds, an
equal number of D6-branes is stretched between every neighbouring pair of
NS-branes. The D6-branes can now reconnect to $K$ infinite D6-branes wrapped
around $x^6$. The system now has a smooth embedding in M-theory as the theory
of
$P$ $E_8\times E_8$ small instantons living at a singularity of type $Z_K$.

The matter content of \asptwo\ was determined in \intr\ by combining arguments
in \am\ with the anomaly cancellation. We see that here it is automatically provided
 by the brane configuration. The case $P<2K$ is obtained by
replacing $K$ with the integer part of $P/2$.
For $P<K$ there is no gauge group. This can be understood from the following
discussion.

As was noted in \am, the matter content in eq. \asp\ is the same as is derived from calculating the three-dimensional mirror of $U(K)$ gauge
theory coupled to $P$ flavours \hw. By taking T-duality three times along the
456
directions, we get the configuration studied in \hw. (To be precise an
additional
mirror symmetry as defined in \hw\ is needed.)
Thus the brane configuration demonstrates that the two effects are indeed
related.
The matter content was calculated using the
rule that an ``S-configuration" is not supersymmetric. See \hw\ for details.
Here we see that this rule is translated into the charge-conservation condition
\cons\ coming from the value of the cosmological constant.

It is now clear why $P<K$ contains no gauge group.

\subsec{More about instantons and enhanced global symmetry}

\lref\ms{D. R. Morrison and N. Seiberg, {\it  Extremal Transitions and
Five-Dimensional Supersymmetric Field Theories}, Nucl. Phys. B483 (1997) 229,
hep-th/9609070}

It is intriguing that the same theories \vec, \wvec\ describe both $SO(32)$ and
$E_8\times E_8$ instantons. The theory \asp\ can indeed be obtained from the
general formulae \ano\ by putting $k=0$ and $W_0=W_P=7,W_K=W_{P-K}=1$. In
general, the global symmetry of the theories \vec, \wvec\ is a subgroup of
$SO(32)$. We just saw that the theory \asp\ has, at the
fixed point,  a global symmetry $E_8\times E_8$, which is not manifest in the
brane picture. It is obtained with a configuration of D8-branes for which the
coupling constant diverges at the orientifold points. By the same mechanism we
can obtain theories with different enhanced symmetry.

The behaviour of the dilaton is determined by the distribution of D8-branes on
the segment. The dilaton is constant only if we put eight branes on each of the 
orientifold points. The background that gives rise to the theory \asp\
has the following behaviour for the dilaton:
\eqn\dil{e^{-\phi}(x^6)=\eqalign{&e^{-\phi_0}+2x^6,\,\,\, 0\le x^6\le x^6_0\cr
&e^{-\phi}(x^6_0),\,\,\, x^6_0\le x^6\le R_6-x^6_0\cr
&e^{-\phi}(x^6_0)-2x^6+2(R_6-x^6_0),\,\,\,
R_6-x^6_0\le x^6\le R_6,}}
where $x^6_0$ is the position of the first D8-brane, $e^{-\phi_0}$ is the value
of the Type I$'$ coupling constant. When $e^{-\phi_0}$ is zero, the dilaton
diverges at both the orientifold points. In general, the dilaton is a linear
function of the position in $x^6$ with
discontinuous derivative at the positions of the D8-branes. The slope of the dilaton for a particular $x^6$ is given by minus twice the value of the cosmological constant 
at that point. 

 We have
an enhanced global symmetry every time the Type I$'$ coupling diverges at the
orientifold point. These configurations are discussed in \seibfive. The basic
building block is a set of $N$ D8-branes on one of the orientifolds. 
In the Type I$'$ strong coupling limit, $SO(N)$ is enhanced to $E_{N+1}$,
where $E_5=$spin$(10),E_4=SU(5),E_3=SU(3)\times SU(2), E_2=SU(2)\times U(1)$
and $E_1=SU(2)$.
There are two additional symmetries, $E_0$ and $\tilde E_1$, coming from a more
careful inspection \ms.

By choosing the positions of the D8-branes we can make the dilaton divergent at
one or both of the orientifold points. For such configurations, we expect to 
find six-dimensional theories with corresponding hidden global symmetry.
The case in which there is an enhancement of symmetry at both orientifolds
with a global symmetry $E_{N+1}\times E_{N^{\prime}+1}$ times
$U(1)$
factors is obtained by putting $N$ D8-branes on the first orientifold and $N^{\prime}$ on the second. The low-energy gauge theory is $\prod_{i=1}^{P-1} SU(V_i)$ with
\eqn\gensmall{\eqalign{V_i= &(8-N)i,\,\,\, 0=k_0<i\le k_1,\, ...\cr
V_i= &V_{k_a}+(8-N-a)(i-k_a),\,\,\, k_a<i\le k_{a+1},a=1,...,15-N-N^{\prime},\cr V_i= &(8-N^{\prime})(P-i),\,\,\,
k_{16-N-N^{\prime}}<i< k_{17-N-N^{\prime}}=P.}}
with bifundamentals for each pair of neighbouring gauge factors and
fundamentals for each factor with $i=k_a$ for some $a\ne 0, 17-N-N^{\prime}$.
The $k_a$'s give the number of NS-branes living on the left of the $a$-th D8-brane (excluding the ones living at the orientifolds). In general, charge conservation gives some constraints on the possible values of $k_a$ for given $N$ and 
$N'$.

One of the properties of the small instantons described in \asp\ is that they leave
an unbroken $E_8\times E_8$ spacetime
gauge group in the vanishing size limit. The breaking of $E_8$ can be obtained by moving some of the D8-branes away from
the
orientifold. We therefore propose that the theories \gensmall, which have
global symmetry $E_{N+1}\times E_{N^{\prime}+1}$, describe the
world-volume theories of
small instantons, which partially break $E_8\times E_8$ to  $E_{N+1}\times E_{N^{\prime}+1}$. 
By arranging the D8-branes on the segment we can 
obtain the different patterns of breaking of the spacetime group by the small
instantons. 
The maximum
value $K$ of $V_i$ in \gensmall\ gives the order of the spacetime singularity $Z_K$ on which
the instantons are sitting. The numbers $k_a$ specify the way in which the
spacetime gauge group is broken by a non-trivial flat connection (which is a
representation of $Z_K$) at infinity.

If some of the D8-branes coincide, there are extra unbroken non-Abelian factors.  For example, small instantons that break the group to
$(E_7\times SU(2))^2$ are obtained by a configuration with 6 D8-branes at each
orientifold and two pairs of coinciding D8-branes on the segment. This
predicts, for $P\ge K$ and $K$ even, a world-volume gauge theory
\eqn\eseven{SU(2)\times SU(4)\times ...\times SU(K-2)\times SU(K)^{(P-K+1)}
\times SU(K-2)\times ... \times SU(2),}
with bifundamentals for each pair of neighbouring gauge factors
and
two fundamentals for each of the first and last $U(K)$ factor coming from the
two pairs of coinciding  D8-branes. 
 The case $P<K$ or $K$ odd can be discussed in
a similar way.

Let us look at the points with enhanced $E_{n+1}$ symmetry from a six-dimensional point of view, and compare them with the points with enhanced
$SO(n)$ symmetry. In the string theory setup, going from $SO(n)$ symmetry to
$E_{n+1}$ requires the limit in which the string coupling is infinite at one of
the orientifold planes. In this paragraph we will call the models with an $SO$
symmetry, $SO$-type theories, and models with an $E$ global symmetry, $E$-type
models.

How does this limit act on the other groups in the models we discuss?
We should recall here that there are two types of non-Abelian groups other than the
ones arising from the D8-branes. These are the gauge theories on the
world-volume of the D6-branes and the groups associated with coinciding NS-branes, which act on strings at the coincident points.

For the gauge theories, by looking at eq. \gaco, we find that all the
gauge couplings diverge. In contrast, for the $SO$
symmetry it is not necessary to flow to the point with infinite gauge coupling
in order to get an
enhanced symmetry; it is just there by tuning the $x^6$ positions of the D8
branes to sit at the orientifold plane\foot{The role of these positions is not
clear from a field theory point of view; however,  we will not discuss them here, since it has been done elsewhere; see, for example, \AH.}.
There are still many non-zero ratios that remain in the process. By observing
eq. \gaco, we see that the ratio between any two different gauge couplings
remains constant in the process of sending the string coupling to infinity.
For the six-dimensional theory these non-zero ratios correspond to the ratios
between various string tensions remaining in the limit. We thus see that the
$E$-type theories have a rich spectrum, which depends on these non-zero ratios.
In contrast, the $SO$ theories  do not have this rich spectrum. At the fixed
point, when all the inverse gauge couplngs are sent to zero, there are no
non-trivial ratios kept in the limit.

Can we see further differences? The tensor multiplets sit in a box with finite
length given in eq. \lengcox. When we take the string coupling to infinity,
the length of the box goes to zero. A crucial difference with the $SO$ models
is that the length of the box is kept finite in the limit of infinite gauge
couplings.

We can now summarize the observations from a six-dimensional point of view.
When the length of the box is finite, the theory admits a global $SO$ symmetry.
The theory flows to a fixed point
when the gauge couplings are taken to infinity.
If in addition we send the length of the box to zero, the global symmetry
is enhanced from $SO$ to $E$. The gauge couplings are
driven to infinity, but we still have the freedom to control their ratio.
The resulting theory will depend on these different ratios, which are 
between various tensions of strings present at the resulting theory.

The matter content in \gensmall\ coincide with that of the three-dimensional mirror of the theory
 \eqn\mir{U(k_1)\times U(k_2)\times ...\times U(K)\times
...\times U(P-k_{15-N-N^{\prime}})\times
U(P-k_{16-N-N^{\prime}})}
with $P$ flavours for the central factor $U(K)$. An explicit
T-duality along the directions 456 transforms the configuration of branes, which
gives rise to the theory \gensmall\ into the three-dimensional mirror of \mir.

The class of theories with $k=0$ in \vec, \wvec\ cannot be used for representing $SO(32)$ instantons, having zero instanton number, but, as we saw, has a natural interpretation
as describing $E_8\times E_8$ instantons. It must be noted that the
interpretation as $SO(32)$ or $E_8\times E_8$ instantons makes sense in a
completely different regime for the Type I$'$ theory. The interpretation of the
system as describing small $SO(32)$ instantons is appropriate for
a generic configuration of background branes in the Type I$'$ theory at weak
coupling, while the interpretation as $E_8\times E_8$ instantons can be
recovered only in the strong coupling limit, when we tune the background brane
configuration in order to get enhanced spacetime gauge symmetry.
As we saw, D6- and
NS-branes interchange their role in representing spacetime orbifold
singularities in the exchange of the two  heterotic string theories. 

\newsec{Six-dimensional string theories}

We now turn to a missing point, the demonstration that the theories considered in the previous sections can be embedded in a consistent six-dimensional string theory
without
gravity \intr. We have to show that the brane system can be embedded in a
well-defined
string theory where its parameters depend on the string length, but not on the
string coupling constant.
In this way, we can send the coupling constant to zero, thus decoupling gravity,
while keeping $l_s$ fixed.

The only parameter in our theories is the size of the Coxeter box. We see
from eq. \lengcox\ that it explicitly depends on the string coupling
constant.
The limit in which the Type I$'$ coupling goes to zero has a Coxeter box of length of order $R_6/l_s^3g_s$; it is therefore not useful for decoupling gravity in our theories. We could keep the Coxeter box size
fixed by sending $R_6$ to zero, but in this limit we should rather do a T-duality and represent the string theory as a type I.

Consider the theories \vec, \wvec. The fact that upon duality
the brane system describes heterotic $SO(32)$ or $E_8\times E_8$ instantons
can be of help. When converted in heterotic variables, the Coxeter box size
indeed becomes of order $l_s^{-2}$. At this point we can send the heterotic string
coupling to zero, decouple gravity and obtain a consistent six-dimensional
theory with scale $l_s^{-2}$.

The infrared limit of these theories, for energies much smaller than $l_s^{-1}$,
is a local quantum field theory. In most cases, we get the non-trivial
interacting fixed points that we discussed in section \nontriv. At higher energies, new
ingredients are needed to give sense to these non-renormalizable theories.
The gravity is decoupled, but the six-dimensional theories still have string
excitations, which appear at a scale of order $l_s^{-1}$. In our brane
construction, they appear as membranes stretched from the NS-branes to their
images under the
orientifolds. These strings have a tension of order $R_6/l_s^3g_s$ in Type I$'$
units, which is
exactly $l_s^{-2}$ in the relevant heterotic variables. Besides having
string-like excitations, these theories inherit a kind of T-duality from the ten-dimensional string
theory in which they are embedded. Therefore, they are not
local quantum field theories.
\lref\lastw{E. Witten, {\it Toroidal Compactification Without Vector Structure}, hep-th/9712028.}
\lref\reykim{N. Kim and S.J. Rey, {\it  Non-Orientable M(atrix) Theory}, hep-th/9710192.}
\lref\dab{A. Dabholkar and J. Park, {\it Strings on Orientifolds}  Nucl. Phys. B477 (1996) 701,  hep-th/9604178.}
For the theories considered in section \generalisation, a dual heterotic model
is more difficult to find. However, we still think that there exists a dual model in which gravity can be decoupled, providing a six-dimensional string theory, which reduces to them at low energies. The orientifold background in which
the theories \veci, \wveci\ are embedded is dual, at strong coupling, to M-theory  compactified on a Klein bottle \refs{\dab,\reykim} and can also be T-dualized to a Type I model
using the results in \lastw. This can provide the way to prove the existence of the associated six-dimensional string theory, but we did not exploited
the subject in detail .

\centerline{\bf Acknowledgements}

We would like to thank Ofer Aharony, Jan de Boer, Erik Gimon, Nissan Itzhaki
and Barak Kol for useful discussions. A.H. would like to thank the CERN
theory division and Instit\"ut f\"ur Physik at Humboldt Universit\"at, for their
hospitality at the final stages of this work.

The research of A.H. is supported in part by NSF grant PHY-9513835.

\listrefs
\end